\documentclass[useAMS]{mn2e}
\usepackage{graphicx}
\usepackage{times}
\usepackage{float}
\usepackage{epsfig}
\newcommand{\bmth}[1]{\mbox{\boldmath${#1}$}}

\title[Dynamic tides in rotating  stars]{
A new general normal mode approach to dynamic tides in rotating
stars with realistic structure and its applications}
\author[P. B. Ivanov, J. C. B. Papaloizou, S. V. Chernov]{P.B.Ivanov$^{1,2}$\thanks{E-mail:
pbi20@cam.ac.uk (PBI)}, J. C. B. Papaloizou $^{2}$\thanks{E-mail:
J.C.B.Papaloizou@damtp.cam.ac.uk (JCBP)}  and S. V. Chernov
$^{1}$\thanks{E-mail: chernov@td.lpi.ru (SVCh)}\\
$^{1}$Astro Space Centre, P.N. Lebedev Physical Institute, 84/32
Profsoyuznaya Street, Moscow, 117997, Russia  \\
$^{2}$ DAMTP, Centre for Mathematical Sciences, University of
Cambridge, Wilberforce Road, Cambridge CB3 0WA }

\begin{document}

\date{Accepted. Received; in original form}

\pagerange{\pageref{firstpage}--\pageref{lastpage}} \pubyear{2010}

\maketitle

\label{firstpage}

\begin{abstract}
We review our recent results on a unified normal mode approach to
dynamic tides proposed in Ivanov, Papaloizou $\&$ Chernov
(2013) and Chernov, Papaloizou $\&$ Ivanov (2013). Our formalism
can be used whenever the tidal interactions are mainly determined
by normal modes of a star with identifiable regular spectrum of
low frequency modes. We provide in the text basic expressions for
tidal energy and angular momentum transfer valid both for periodic
and parabolic orbits, and different assumptions about efficiency
of normal mode damping due to viscosity and/or non-linear effects
and discuss  applications to binary stars and close orbiting
extrasolar planets.
\end{abstract}

\begin{keywords}
hydrodynamics - celestial mechanics - planetary systems:
formation, planet -star interactions, stars: binaries: close,
rotation, oscillations, solar-type
\end{keywords}

\section{Introduction}

In this contribution we briefly discuss our results on a unified
normal model approach to dynamic tides and its applications
published in Ivanov, Papaloizou $\&$ Chernov (2013) and Chernov,
Papaloizou $\&$ Ivanov (2013) hereafter Paper 1 and 2,
respectively. Our formalism is applicable when the tidal
interactions are mainly determined by normal modes of a star with
identifiable regular spectrum of low frequency modes, such as
rotationally modified gravity modes. In this case it allows one to
consider the well known  theory of dynamic tides (e.g. Zahn
1977, Goodman $\&$ Dickson 1998) operating in binaries with
 periodic orbits as well as the  excitation of normal modes  as the result of  a
 flyby around  a perturbing body by a rotating star
moving on  a parabolic orbit as discussed by Press $\&$ Teukolsky
1977 (see also, e.g. Lai 1997, Ivanov $\&$ Papaloizou 2004, 2007,
hereafter IP7).  These are considered in a general way in parallel
developments.

It is shown that in general,  energy and angular momentum transfer
due to dynamic tides is  determined by the so-called overlap
integrals provided that a mechanism of dissipation of energy
stored in normal modes is specified and orbital parameters
together with the state of rotation of the star are given. It is
stressed that in certain important cases,   such as  dynamic tides
operating in the Press $\&$ Teukolsky sense, or dynamic tides
operating in the Zahn sense in the regime of moderately large
dissipation, for which internally excited waves travelling with
their group velocity do not reach either the centre or boundary of
the configuration on account of damping, the resulting expressions
for either the energy and angular momentum transfer rates, or the
total amount transferred in each case, do not depend on details of
energy dissipation, see Section \ref{s1}.

In Section \ref{s2} we briefly  present the overlap integrals
obtained  for Sun-like stars and stars with mass
$M_*=1.5M_{\odot}$ referring the reader to Paper 1 and Paper 2,
where these quantities are calculated numerically for a range of
rotating Population I stellar models of different masses and ages.
These results are obtained for the slow rotation regime where
centrifugal distortion is neglected in the equilibrium and the
traditional approximation is made, and an analytical WKBJ theory
of their evaluation is developed for the case of Sun-like rotating
stars.

In Section \ref{s3} we use  our formalism to  determine the tidal
evolution time scales for an  object,  in an  orbit of small
eccentricity, around a Sun-like star in which the tidal response
is assumed to occur. This is taken to have a perturbing companion,
either with the same mass, or with a mass typical of exoplanets.
We consider the case of synchronous  rotation and in addition a
non rotating star when the companion is an exoplanet.  We assume
that wave dissipation occurs close to the stellar centre such that
the moderately large dissipation regime applies.
 Additionally, we apply the formalism
to the problem of tidal excitation of normal modes after a
periaston flyby showing that realistic stars having an extended
convective envelope are more susceptible to the action of tides
than a frequently used reference model consisting of a polytrope,
with index $n=3,$ and  with  the same mean density, see Section
\ref{s4}.

\section{General expressions for the energy and angular momentum
transfer} \label{s1}

\subsection{Basic definitions}

We assume below that when  the orbital motion may be considered to be
periodic a discrete Fourier series in   time is appropriate.  For
a real quantity, $Q,$  we write
\begin{equation}
Q = \sum_{m, k} \left( Q_{m,k}\exp({-{\rm i}\omega_{m,k}
t+im\phi}) + cc \right ), \label{eq p1}
\end{equation} where $cc$ denotes the complex conjugate,
$\omega_{m,k}=k\Omega_{orb}-m\Omega$, $\phi$ is the azimuthal
angle in the spherical coordinate system $(r, \theta, \phi)$
associated with a frame rotating with the angular frequency
$\Omega $ with respect to an inertial frame with the rotation
axis being directed along $z=r\cos \theta$.  The
orbital frequency  is $\Omega_{orb}$,  and $m$ and $k$ are integers, with
only positive values of $m$ included in the summation as in IP7.
When the orbital motion may be treated as parabolic we employ  a
Fourier transform instead of a  Fourier series. The stellar
perturbations are described by the Lagrangian displacement vector
${\bmth{ \xi}}$. The  dominant quadrupole component of the tidal
potential, $\Psi $, is represented in the form
\begin{equation}
\Psi=r^{2}\sum_{m,k} \left( A_{m,k}e^{-i\omega_{m,k}t} Y^{m}_{2} +
cc \right) \label{eq4}
\end{equation}
where $Y^{m}_{2}$  are the spherical functions. The coefficients
$A_{m,k}$ are given in Appendix A of Paper 1 for the case of
coplanar orbit with small eccentricity. For orbits having an
arbitrary value for the  eccentricity,  these coefficients can be
expressed in terms of so-called Hansen coefficients discussed eg.
in Witte $\&$ Savonije (1999).  We use a parallel development
formulated in terms of the Fourier transform of the tidal
potential in the case of parabolic orbits.

\subsection{The energy and angular momentum transfer rates in the case
of periodic orbits}

In this case the general expressions are obtained in Paper 1. The
rate of transfer of energy in the rotating frame, $\dot E_c$, the
rate of transfer of angular momentum, $\dot L_c$, and the rate of
transfer of energy in the inertial frame,  $\dot E_I$, read as
\begin{eqnarray}
\dot E_c&=&-\sum_{m,k,j}{\omega_{\nu,kj}
{|A_{m,k}\hat Q_{j}|^2\over  D_{k,j,j}}}, \nonumber \\
\dot L_c&=&- \sum_{m,k,j}m{\omega_{\nu,kj}\over \omega_{m,k}}{
|A_{m,k}\hat
Q_j|^2\over D_{k,j}}   {\hspace{1cm}} {\rm and}\nonumber \\
  \dot E_I&=&-
\sum_{m,k,j}{\omega_{\nu,kj}(\omega_{m,k}+m\Omega)\over
\omega_{m,k}}{ |A_{m,k}\hat Q_j|^2\over D_{k,j}}. \label{eq33}
\end{eqnarray}
Here
\begin{equation}
D_{k,j}=(\omega_{m,k}-\omega_{j})^2+\omega_{\nu,kj}^2,
\label{eq23}
\end{equation}
where $\omega_j$ are eigenfrequencies of normal modes with the
index $(j)$ designating particular modes, and $\omega_{\nu,kj}$
are their damping rates.

The overlap integrals are
\begin{equation} \hat Q_j=Q_j/\sqrt{n_j},
\label{eq33a}
\end{equation}
where
\begin{equation}
Q_{j}=\int d^3x\rho e^{-im\phi}{\bmth \xi}^{*}_{j}\cdot \nabla
(r^{2}Y^{m}_{2}). \label{eq31}
\end{equation}
and
\begin{equation}
n_{j}=\pi (({\bmth {\xi}}_{j}| {\bmth {\xi}}_{j})+({\bmth
{\xi}}_{j}|{\bmth {\cal C}}{\bmth{\xi}}_{j})/\omega_{j}^2),
\label{eq32}
\end{equation}
and ${\cal C}$ is an integro-differential operator describing the
action of pressure and gravity forces on stellar perturbations.
Here dot and asterisk stand for the scalar product and complex
conjugate, respectively.

\subsubsection{A dense spectrum of modes}

Let us assume that the  difference between neighboring values of
eigenfrequencies is much smaller than their  typical absolute values
 and that for a particular forcing
frequency, $\omega_{m,k},$ the
 resonance condition is most closely satisfied for a particular
mode with  index $j=j_{0},$ (which we recall will vary with $m$
and  $k$). Accordingly,  we have
\begin{equation}
\omega_{j_{0}}=\omega_{m,k}+\Delta \omega_{j_{0}}, \label{eq26}
\end{equation}
where   the offset, $\Delta \omega_{j_{0}}, $ is such that
 $|\Delta \omega_{j_{0}}| < |{d\omega_{j_{0}} /
dj_{0}}|$. In this case the general expressions (\ref{eq33}) can
be simplified. In general, their form depends significantly on two
dimensionless parameters

They take  an especially simple form depending on whether the
parameter $\kappa =|\omega_{\nu,kj_0}/ d\omega _{j0} /dj_0|$ that
measures the effectiveness of dissipation
 is large or small. As long as $|\omega_{\nu,kj_0}/\omega _{j0}| \ll 1,$
the former case corresponds to the so-called regime of moderately
large  dissipation,  for which the decay time due to viscosity is
much smaller than the time for  waves  to propagate across the
star with their group velocity, see e.g. Goodman $\&$ Dickson
(1998). In this case we have
\begin{eqnarray}
\dot E_{c}&=&-\pi \sum_{m,k}{|A_{m,k}\hat Q_{j_{0}}|^2\over
|{d\omega_{j_{0}}/ dj_0}|},  \nonumber \\
 \dot L_{c}&=&-\pi \sum_{m,k}{m\over
\omega_{m,k}}{|A_{m,k}\hat Q_{j_{0}}|^2\over |{d\omega_{j_{0}}/
dj_0}|}, \hspace{1cm}{\rm and}   \nonumber \\
\dot E_{I}&=&-\pi \sum_{m,k}\left(1+{m\Omega\over
\omega_{m,k}}\right ){|A_{m,k}\hat Q_{j_{0}}|^2\over
|{d\omega_{j_{0}}/ dj_0}|}.
 \label{eq34}
\end{eqnarray}
In the latter case when  viscosity is very small, and,
accordingly, $\kappa \ll 1$, we obtain
\begin{eqnarray}
\dot E_c&=&-\sum_{m,k}{\omega_{\nu,kj_0}\over D_*}
|A_{m,k}Q_{j_{0}}|^2, \nonumber \\
\dot L_c &=&-\sum_{m,k}{m\omega_{\nu,kj_0}\over \omega_{m,k}D_*}
|A_{m,k}Q_{j_{0}}|^2,  \hspace{1cm}  {\rm and} \nonumber   \\
\dot E_I&=&-\sum_{m,k}\left (1+{m\Omega\over
\omega_{m,k}}\right){\omega_{\nu,kj_0}\over D_*}
|A_{m,k}Q_{j_{0}}|^2, \label{eq35}
\end{eqnarray}
where $D_{*}=(\omega^2_{\nu,kj_0}+ ({d\omega_{j_{0}}/ dj_0})^2\sin
^2 \pi \delta/ \pi^2),$

\noindent and $\delta = |\Delta \omega_{j_0}/ d\omega _{j0}
/dj_0|.$

\subsection{Energy and angular momentum transfer in the case of
a prescribed parabolic orbit}

The transfers of energy, $\Delta E_{c}$, $\Delta E_{I}$ and
angular momentum, $\Delta L_{c}$, between the orbital motion and
normal modes of the star during a periastron flyby on a prescribed
parabolic orbit can also be expressed in terms of overlap
integrals. For completeness we give expressions for these here. As
mentioned above, in this case we should represent the results in
terms of the Fourier transform of the tidal potential instead of
the Fourier  coefficients used in the previous sections.

Expressions for $\Delta E_{c}$, $\Delta E_{I}$ and $\Delta L_{c}$
in terms of this were derived by IP7 \footnote{ Note that  in
equation (5) of IP7 there should be a $(-)$  sign in front of the
expression on the right hand side and  in equation (28) the
imaginary unit should  be inside the summation.}. As in the case
discussed above they also can be represented as sums over
quantities determined by the eigenmodes, which for a given $m$
have eigenfrequency $\omega_k,$ where, for convenience,  $m$ is
here suppressed as a suffix. They take  the form
\begin{eqnarray}
\Delta E_{c}&=&{2\pi^2}\sum_{m,k}{|A_{m}(\omega_k)\hat Q_k|^2}, \nonumber \\
\Delta L_{c} &=&{2\pi^2}\sum_{m,k}{m \over \omega_k}
{|A_{m}(\omega_k)\hat Q_k|^2}, \nonumber \\
\Delta E_{I} &=& {2\pi^2}\sum_{m,k}\left (1+ {m\Omega\over
\omega_k}\right ) {|A_{m}(\omega_k)\hat Q_k|^2}. \label{eq37}
\end{eqnarray}
Note that the quantities $A_m(\sigma)$ entering (\ref{eq37}) are
discussed in e.g. Ivanov $\&$ Papaloizou (2011) for the  general
case of a parabolic orbit inclined with respect to equatorial
plane of a rotating star\footnote{The analogous expressions of
Ivanov $\&$ Papaloizou (2011) differ by factor of two from those
given in (\ref{eq37}). This is due to the fact that in Ivanov $\&$
Papaloizou (2011),  the summation over the azimuthal mode number
is formally performed  over positive and negative values of $m$,
while in this paper we consider  only positive azimuthal mode
numbers. Note also that in this paper we take all quantities
related to eigenmodes to be defined in the rotating frame,  while
in Ivanov $\&$ Papaloizou (2011) the inertial frame was used.} .

\section{Overlap integrals} \label{s2}

\begin{figure}
\begin{center}
\vspace{8cm}\includegraphics{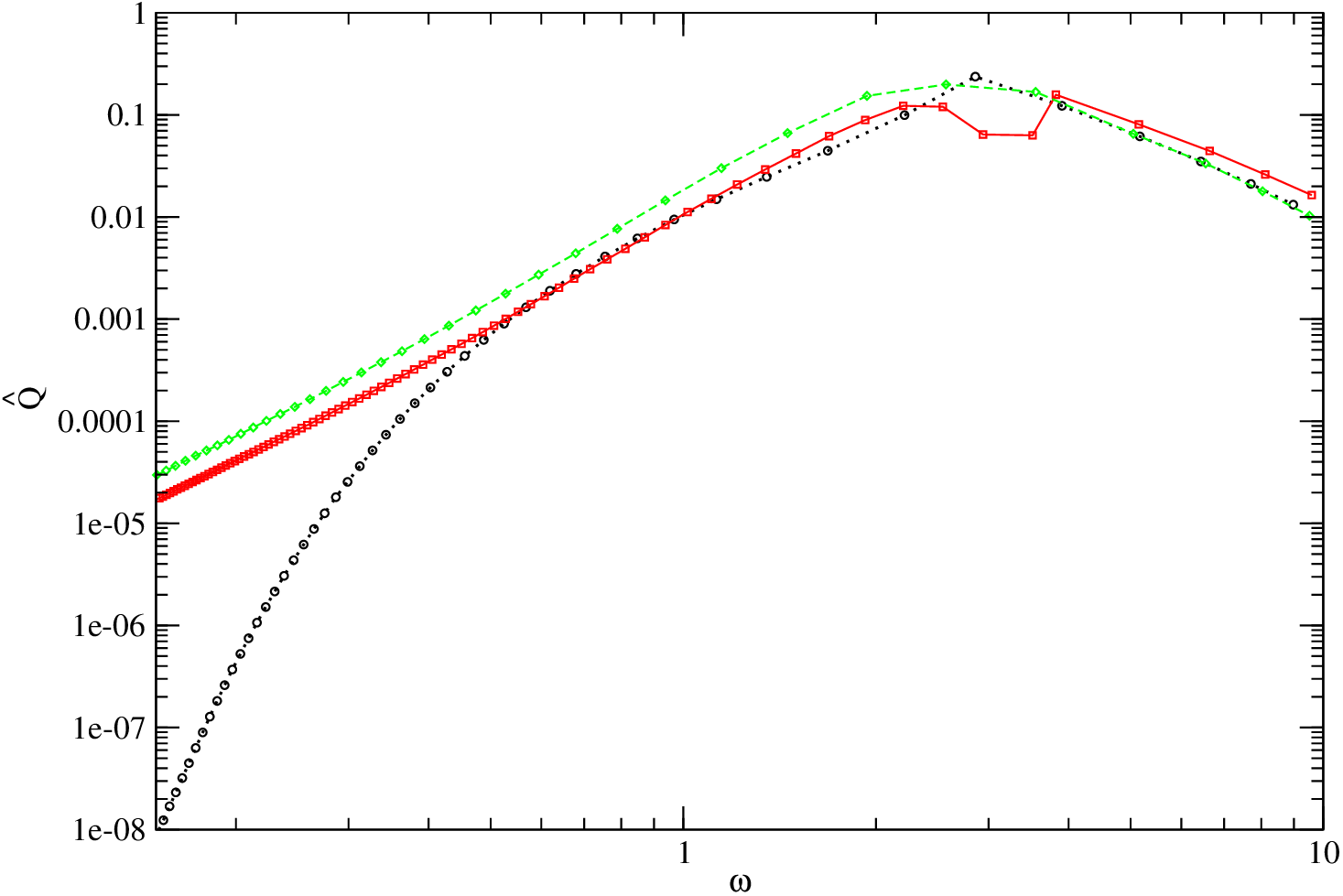}
\end{center}
\vspace{0.5cm} \caption{The overlap integrals $\hat Q$ as
functions of mode eigenfrequency $\omega $ for models 1a and 1b
plotted with  dashed and solid curves, respectively. The dotted
curve plots the overlap integrals for  a polytrope with $n=3.$
Note that for low frequencies,  these are much smaller than the
ones corresponding to stellar models with realistic structure.
Symbols show the positions of eigenfrequencies. The smooth curves
are interpolated through these.} \label{Fig5}
\end{figure}

\begin{figure}
\begin{center}
\vspace{8cm}\includegraphics{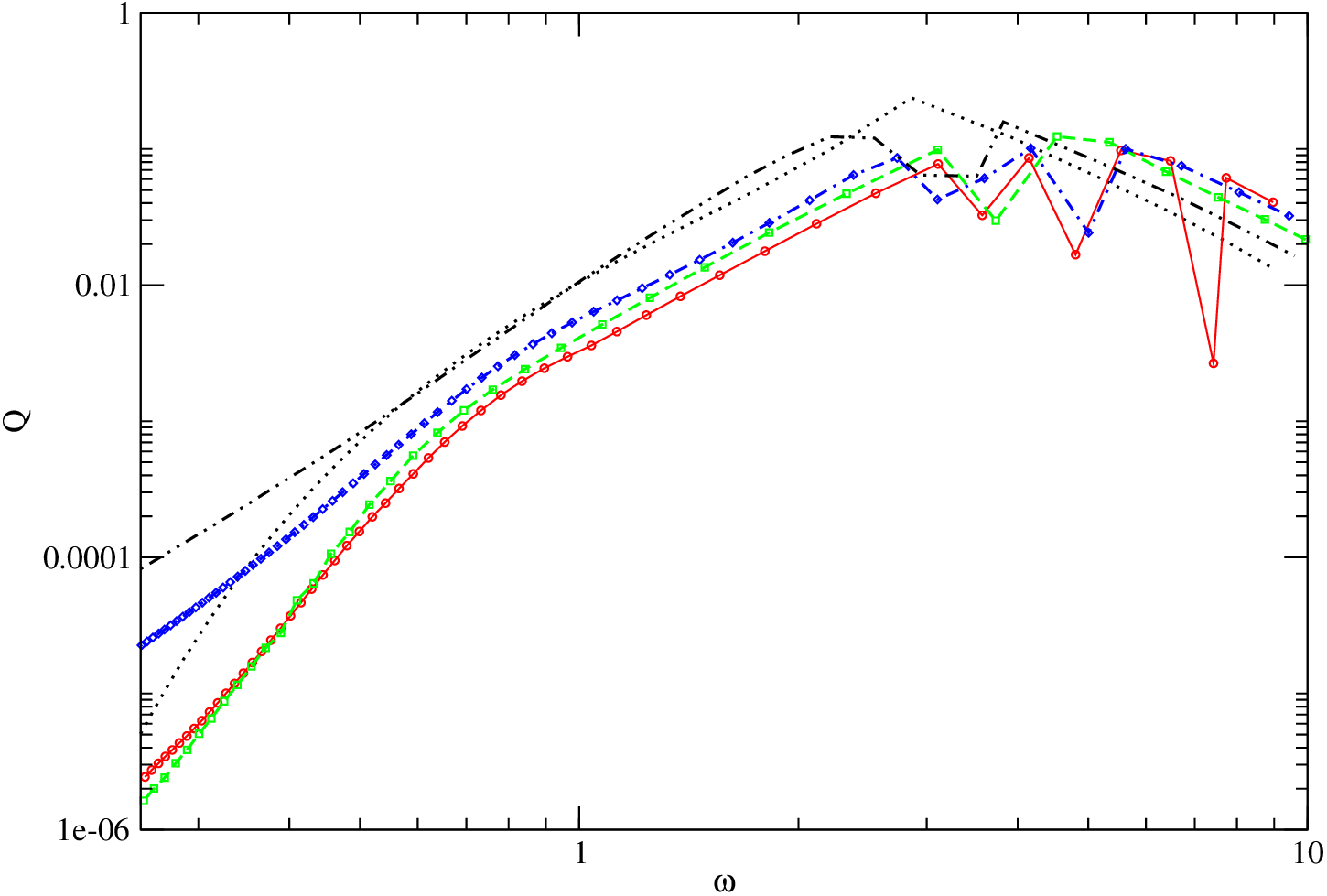}
\end{center}
\vspace{0.5cm} \caption{As for  Fig. \ref{Fig5} but for models
with $M_*=1.5M_{\odot}$. Dot dashed, dashed and solid curves are
for models 1.5a, 1.5b and 1.5c, respectively. We show the results
for models 1p and 1b,  plotted as dotted and dot  dot dashed
curves respectively, for comparison.} \label{Fig6}
\end{figure}

In this Section we present the overlap integrals for two models of
Sun-like stars with mass $1M_{\odot}$, model 1a and 1b having
ages $1.67\times 10^8yr$ and $4.41\times 10^9yr$, respectively, and
three models of a star with mass $1.5M_{\odot}$    with ages
$1.27\times 10^7yr$ (model 1.5a), $5.96\times 10^7yr$ (model 1.5b)
and $1.58\times 10^9yr$ (model 1.5c), see Figs \ref{Fig5} and \ref{Fig6}. 
The stellar models were
calculated by Christensen-Dalsgaard (1996) and Roxburgh (2008). We
also consider a polytrope with index $n=3$ as a reference model, denoting it as
model 1p. The overlap integrals $Q$ are expressed in the natural
units $\sqrt{M_*}R_{*}$, where $M_*$ and $R_*$ are the stellar
mass and radius, respectively, and the eigenfrequencies are
expressed in terms of $\Omega_*=\sqrt{GM_*/ R_*^3}$, where
$G$ is the gravitational constant. All stars are assumed to be
non-rotating. A discussion of effects due to rotation, the
calculation of overlap integrals for stars with larger masses can
be found in Paper 2 and the comparison of the numerical results
shown here with the corresponding analytical expressions for
Sun-like stars can be found in Paper 1. We only mention here that
the comparison shows a good agreement between the two approaches, even
for fast rotators.
One can see from Fig. \ref{Fig5} the Sun-like models have much
larger values of the overlap integrals as compared to those for
the polytropic model at small eigenfrequencies, $\omega $. As
discussed in Paper 1 it is due to the presence of extended
convective envelopes in the solar models, which result in a power
law decay of $\hat Q$ with  $\omega $ in comparison to  an
exponential decay found for  the polytropic model. This means that
tidal interactions of Sun-like stars at sufficiently low forcing
frequencies are stronger than those corresponding to a polytropic
star   with the same mean density, see also Fig. \ref{Fig9} below.
The $\hat Q$ corresponding to models with $M_*=1.5M_{\odot}$ are
smaller than those of Sun-like stars. This is because the
convective envelopes are much less pronounced in the models of
more massive stars discussed in the text. For the shown range of
eigenfrequencies, values of $\hat Q$ for more massive models are
even smaller than  those corresponding to the polytropic model
with the exception of the youngest model 1.5a, which has indeed
the most extended convective envelope among the models with
$M_*=1.5M_{\odot}$ discussed here.

\section{Application of the formalism to the case of a binary orbit with  small eccentricity} \label{s3}

\begin{figure}
\begin{center}
\vspace{8cm}\includegraphics{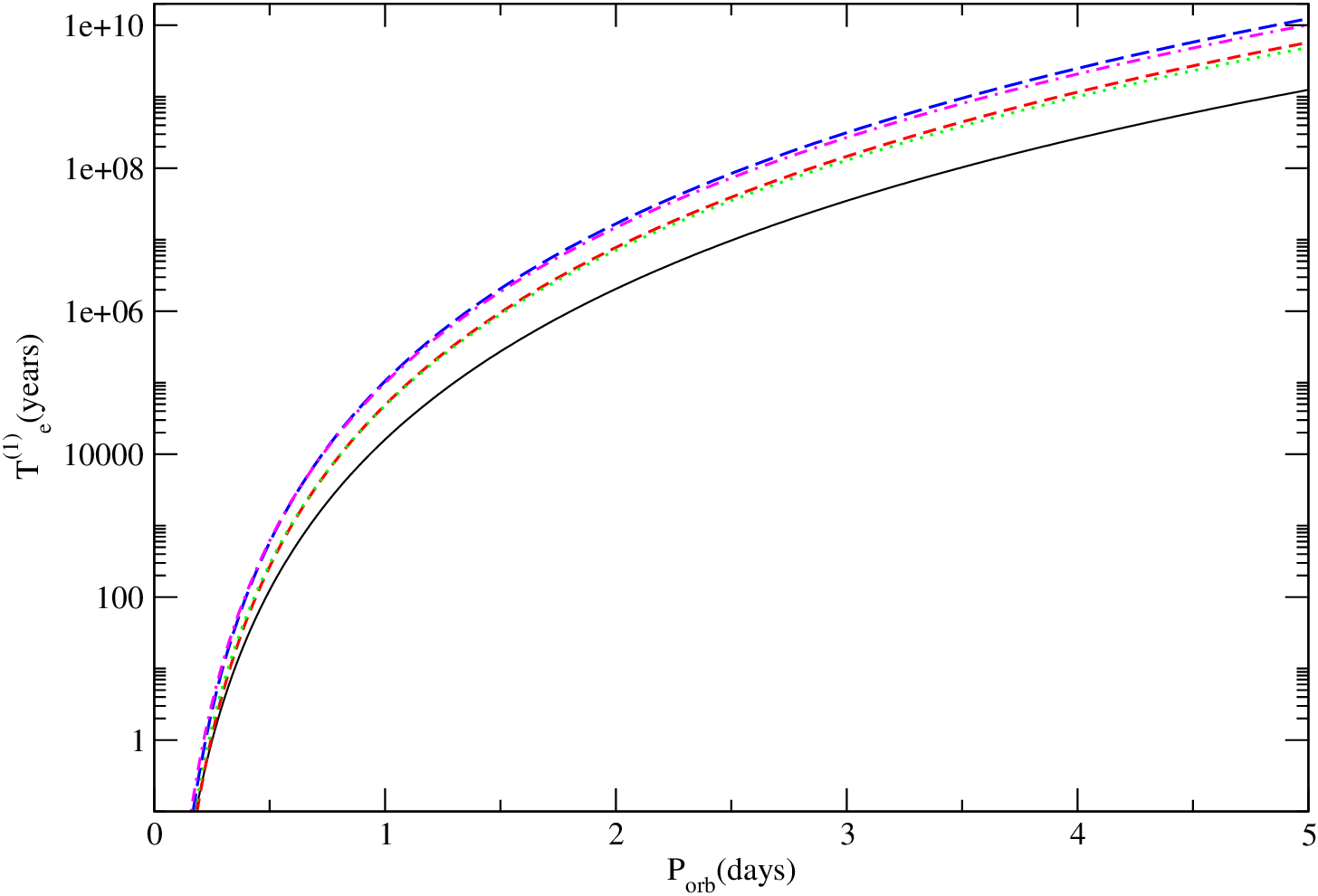}
\end{center}
\vspace{0.5cm} \caption{The time scale for evolution of
eccentricity, $T^{(1)}_{e}$, calculated for the 1a and 1b models
compared to the result of Goodman $\&$ Dickson (1998), for
$\mu=1$. See the text for the description of different curves.}
\label{Fig10}
\end{figure}

In order to illustrate applications of our formalism let use
discuss two simple problems assuming that the regime of moderately
large viscosity operates in the star where tides are raised. The
first concerns a binary consisting of a Sun-like star in a state
of synchronous rotation and a point-like object of the same mass
in an orbit of small eccentricity, $e$. We plot the
corresponding time scale for the decay of eccentricity in Fig.
\ref{Fig10}.
The short dashed and dotted curves are for model 1b with curves of
different types corresponding to different fits for  the Brunt -
V$\ddot {\rm a}$is$\ddot {\rm a}$l$\ddot {\rm a}$ frequency close
to the base of convection zone while the long dashed and dot
dashed curves are for model 1a. One can see from Fig. \ref{Fig10}
that there is little difference between the curves corresponding
to different fits. The solid line plots  a minor
modification of the expression of Goodman $\&$ Dickson (1998)  who
give $T_{e}^{GD}=8\cdot 10^3P_{orb}^7,$ with $P_{orb}$ in days,
for a system of two tidally interacting Sun-like stars of equal
mass. Since we assume that the tides are raised only on one
component, we plot $T_{e}^{GD}=1.6\cdot 10^4P_{orb}^7$ in Fig.
\ref{Fig10}. Our results are seen to give larger values of $T_{e}$
for a given $P_{orb},$ e.g.  for $P_{orb} = 4 $ days, the results
differ by  a factor of five to ten.  Thus we have $T_{e}\approx
2.7\cdot 10^8yr$, $1.15\cdot 10^9yr$ and $2.35\cdot 10^9yr$ from
the modified  Goodman $\&$ Dickson (1998) expression and for our
calculations corresponding to models 1b and 1a, respectively.

\begin{figure}
\begin{center}
\vspace{8cm}\includegraphics{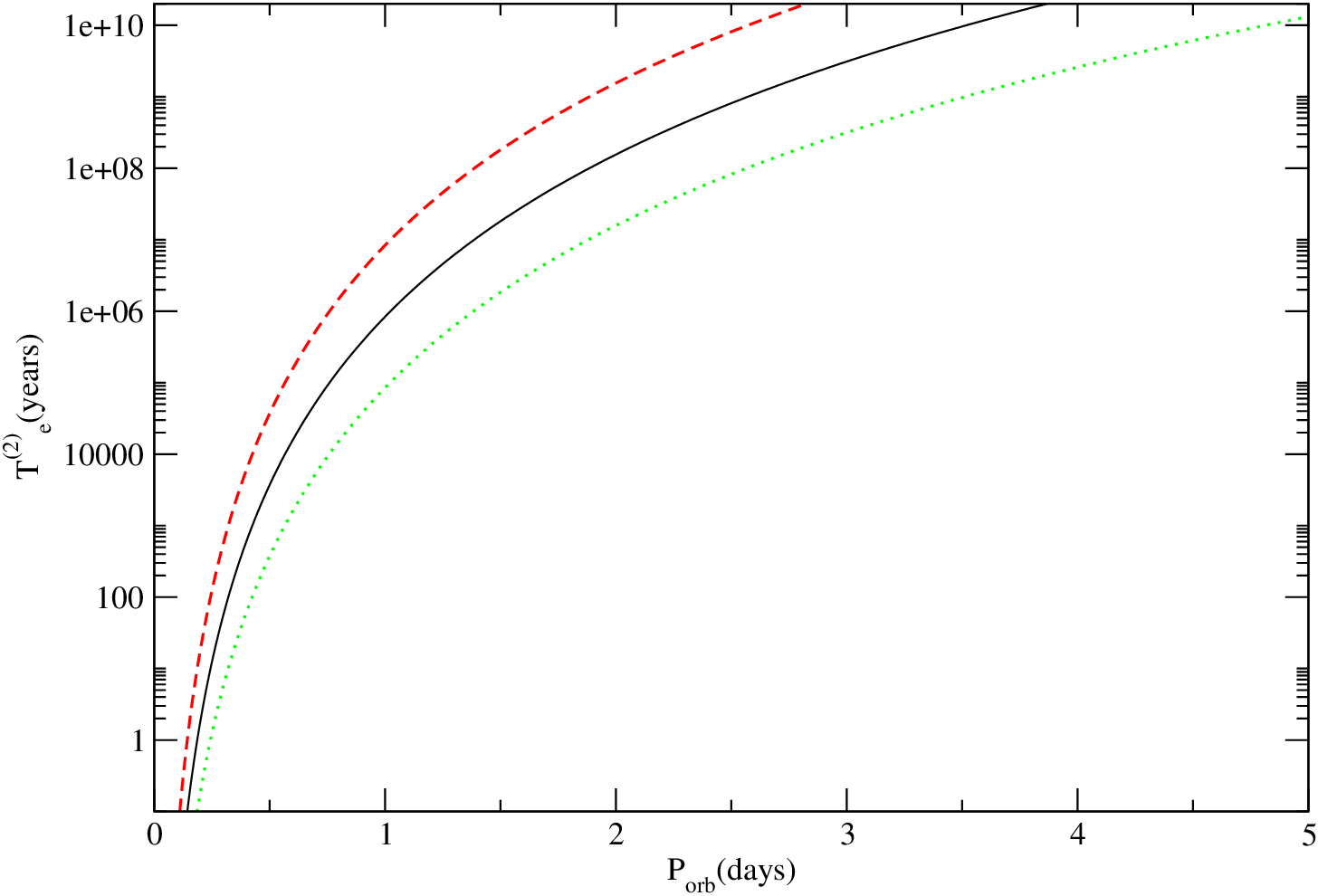}
\end{center}
\vspace{0.5cm} \caption{The evolution time scales for orbital
eccentricity, $T^{(2)}_e$,  plotted for the mass of perturbing
object, $m_p=M_{J}$, $0.1M_{J}$ and $10M_{J}$ using the solid,
dashed and dotted curves, respectively. } \label{Fig12}
\end{figure}

\begin{figure}
\begin{center}
\vspace{8cm}\includegraphics{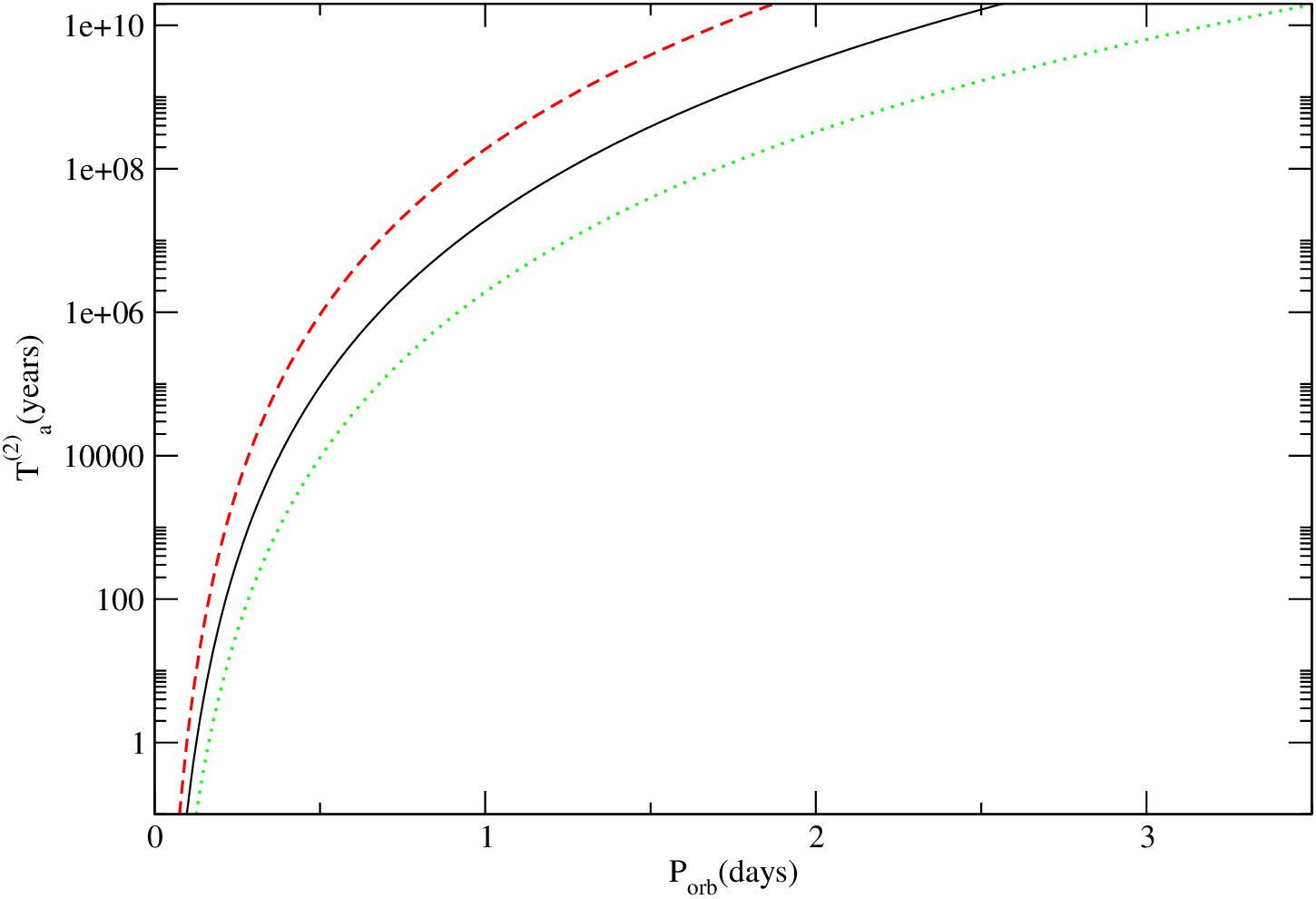}
\end{center}
\vspace{0.5cm} \caption{Same as Fig. \ref{Fig12} but the evolution
time scale of the semi-major axis, $T^{(2)}_a$, is shown.}
\label{Fig13}
\end{figure}

The second problem deals with orbital evolution of a point-like
object with a mass appropriate for a massive exoplanet orbiting
around a non-rotating Sun-like star. The corresponding time scales
of the evolution of eccentricity and semi-major axis are shown in
Fig. \ref{Fig12} and \ref{Fig13}, respectively.
The solid, dashed and dotted curves there correspond to the
perturbing mass $m_p=M_{J}$, $0.1M_{J}$ and $10M_{J}$. As seen
from these plots, dynamical tides in the regime of moderately
large viscosity are potentially efficient only when orbital
periods are rather small. Thus, the eccentricity is damped in  a
time less than $10^9yrs$ only for $P_{orb} < 1.85$, $2.56$ and
$3.48days$, for $m_p=0.1M_{J}$, $M_{J}$ and $10M_{J}$,
respectively. Similarly the semi-major axis can decay  in a  time
less than $10^9yrs$ only for $P_{orb} < 1.24$, $1.69$ and
$2.3days$ respectively.

\section{Energy transferred as the result of a flyby
 around a centrally condensed
mass on parabolic orbit} \label{s4}

Now let us consider the problem of parabolic flyby of a star
around a point-like source of gravity. One can show that once the
ratio of the rotational frequency to the characteristic stellar
frequency, $\Omega /\Omega_*$, is specified, the energy and
angular momentum transferred, $\Delta E$ and $\Delta L$, expressed
in units of $E_*=Gm^2_p/( (1+q)^2 R_*)$ and $L_*={q^2
(1+q)^{-2}}M_*\sqrt{GM_*R_*}$, where $q=m_p/M_*$, respectively,
are functions of only one parameter (see eg. Press $\&$ Teukolsky
1977, Ivanov $\&$ Papaloizou 2004, 2007)
\begin{equation}
\eta=\sqrt{{1\over 1+q}{\left({R_p\over
R_*}\right)}^3}=3.05\sqrt{\bar \rho}P_{orb} ,\label{eq2}
\end{equation}
where $R_p$ is the periastron distance. The quantity $\bar \rho$
is the ratio of  the mean stellar density to the solar value,
$\bar \rho ={R_{\odot}^3M_* /(R_{*}^{3} M_{\odot})},$ and
$P_{orb}$ is orbital period of a circular orbit which has  the
same value of the  orbital angular momentum as the parabolic orbit
under consideration, expressed in units of one day. Here we
consider only the case of non-rotating Sun-like star referring the
reader to Paper 2 and Ivanov $\&$ Papaloizou 2011, where more
general cases are discussed.

\begin{figure}
\begin{center}
\vspace{8cm}\includegraphics{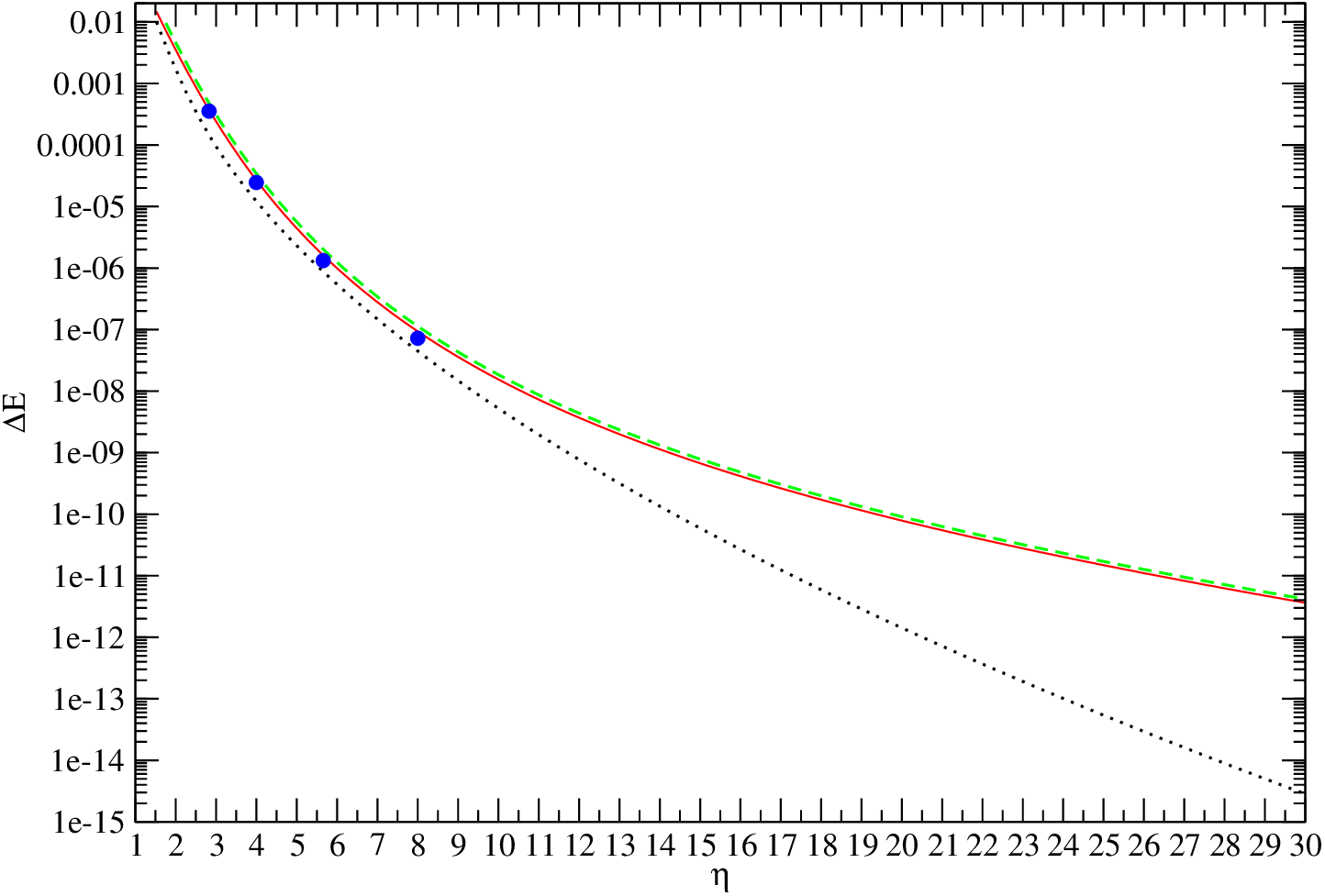}
\end{center}
\vspace{0.5cm} \caption{Energy transferred to the normal modes of
a non-rotating star with $M_*=M_{\odot}$ as the result of  a
parabolic flyby of a perturber of mass $m_p=M_J$, expressed in
units of $E_*$, as a function of the parameter $\eta $. The solid,
dashed, and dotted curves are for models 1b, 1a and 1p,
respectively. The dot dashed curve shows results  calculated using
the  purely analytic expressions for the mode eigenfrequencies and
overlap integrals obtained in Paper 1 for model 1b.  Circles show
the energy transfer calculated for model 1b using the direct
numerical approach.} \label{Fig9}
\end{figure}
The energy transferred as a result of tidal encounter is shown in
Fig. \ref{Fig9}. One can see that the realistic stellar models
give much larger values of $\Delta E$ at sufficiently large values
of $\eta $ as compared with the reference polytropic model.
 This effect is due to the power law dependence of the overlap
integrals in the limit $\omega \rightarrow 0$ mentioned above,
which is, in turn, is due to the presence of convective regions in
realistic stellar models.

\section{Conclusions}

In this contribution we present a new general approach to the
problem of dynamic tides. It allows one, in principal, to
calculate all quantities of interest both for a periodic orbit of
any eccentricity and for a prescribed  parabolic orbit. It
generalizes the well known approaches of Zahn and Press $\&$
Teukolsky. It is shown that provided the rate of
 decay of free stellar oscillations  due to either linear dissipation
or non-linear effects,  as well as the orbital parameters,  are
specified,  all information about tidal interactions is encoded in
the overlap integrals. It is stressed that when either a system is
evolving in the so-called regime of moderately large viscosity or
undergoes a flyby around a gravitating body,  the expressions
governing energy and angular momentum exchange do not depend on
the quantities describing the decay of free modes. The overlap
integrals are calculated numerically for a set of Population I
stellar models of different masses and ages, and, additionally,
analytically for Sun-like stars with convective envelope and
radiative core. The formalism is applied to the orbital evolution
of systems having periodic orbits with small eccentricity evolving
in the regime of moderately large viscosity and the parabolic
flyby problem. In the former case, it is shown that a Hot Jupiter
can spiral in a host Sun-like star in a time smaller than or of
the order of $10^9$ when its orbital period is smaller than $\sim
1.7d$. In the latter case it is shown that, in general, stars with
realistic structure have much stronger tidal response at
sufficiently large values of impact parameters than the reference
polytropic model provided that they have an extended convective
envelope. One  unresolved issue is to provide an extension of the
analytic methods of calculation of the overlap integrals in the
limit $\omega \rightarrow 0$ to the case of stars with convective
cores and radiative exteriors. This is left for a possible future
work.

\section*{Acknowledgments}

We are grateful to I. W. Roxburgh and G. I. Ogilvie for help and
fruitful discussions and to A. A. Lutovinov for useful comments.

PBI and SVCh were supported in part by RFBR
grant 15-02-08476a, by programmes 9 and 22 of the Presidium of the Russian Academy
of Sciences, and by the grant of the President of the Russian Federation
for Support of Leading Scientific Schools of the Russian Federation
NSh-4235.2014.2. Additionally, SVCh is grateful to RFBR grant 14-02-00831.

\end{document}